\documentclass[preprintnumbers,amsmath,amssymb,onecolumn]{revtex4}
\usepackage[dvips,colorlinks=true,citecolor=red,filecolor=green,linkcolor=blue,pdfnewwindow=true]{hyperref}
\usepackage{graphicx}
\usepackage{color}
\usepackage{multirow}
\usepackage[title]{appendix}

\begin{document}
\title{Functional switching among dynamic neuronal hub-nodes in the brain induces transition of cognitive states.}
\author{Jasleen Gund\textsuperscript{1,3}, Yashaswee Mishra\textsuperscript{2}, R.K. Brojen Singh\textsuperscript{1*}\thanks{Corresponding author.}, B.N. Mallick\textsuperscript{2}\thanks{Corresponding author.}}
\email{brojen@jnu.ac.in (Corresponding author)}
\email{bnm@mail.jnu.ac.in (Corresponding author)}
\affiliation{$^1$School of Computational and Integrative Sciences, Jawaharlal Nehru University, New Delhi-110067, India.\\
$^2$School of Life Sciences, Jawaharlal Nehru University, New Delhi-110067, India.\\
$^3$CBDL, National Brain Research Centre, Manesar, Gurgaon-122050, India.}

\begin{abstract}
{\noindent}The cognitive states have broadly been divided into waking, rapid eye movement sleep (REMS) and non-REMS (NREMS). Although the mechanism of state transition is unknown, it has been proposed that functional activation/deactivation among different brain regions leads to such transition. As analysis of electroencephalogram (EEG) allows us exploring properties and association among brain regions, we exploited it to address our query. We have recorded the frontal and occipital cortical EEG from surgically prepared chronic freely moving, normally behaving rats and classified their vigilance states (VS) and vigilance state-transitions (VST). The complexity analysis carried out by computing multifractal spectrum width categorized VST as highly non-linear and complex than their participating vigilance states. The EEG signals were decomposed into frequency ranges as that of classical human Delta (0.5-4 Hz), Theta (4-7.5 Hz), Alpha (8-12 Hz), Beta (13-20 Hz) and Gamma (21-50 Hz) oscillations. The \textit{dynamic network attributes} of these oscillations has shown compelling topological correlation between the frontal and occipital regions of the brain during conscious states. The topological trends of the underlying hierarchical network organization have been characterized by the percentage of the hub and non-hub nodes, that further determines the global and local connectivity trends. This has revealed interesting insights of functional disconnection during NREMS due to decrease in their number of hub-nodes as compared to waking state. There is also switching behaviour in the ratio of hub/non-hub nodes between frontal and occipital region during NREMS-Wake and Wake-NREMS transitions. Our findings provide support as proof-of-principle of functional regional inactivation or activation of hub-nodes as the gradual switching mechanism towards transitioning of cognitive states in a graded (or non-flipping) manner, wake to sleep or vice versa; the detailed neuro-physio-chemical mechanism needs further study.
\end{abstract}
 

\maketitle
\vskip 1cm
{\noindent}\textbf{\Large Introduction}\\

{\noindent}The spontaneous, apparently random electrical signals of the brain, the electroencephalogram (EEG), manifests electrical activities among clusters of neurons in the brain, which may or not be connected and is considered to be a reasonable objective correlate of the dynamic cognitive states. Usually the EEG recording sites are fixed (once the electrodes are connected), however, as the locations of the source are dynamic and may shift, the distance and phase between the source and recording sites vary continuously. As the temporal and spatial inputs on the source neurons to trigger their output vary dynamically, the phase and intensity of the recorded signals vary continuously. The non-synchronized activation or deactivation of neurons within the same or different clusters would induce variation in phase, polarity and intensity of signals at the recording site. This is likely to induce a gradual changes in the potential in the neuronal cluster(s), which are reflected in the EEG. As such the gross EEG waves are highly complex and dynamic, which vary depending on the activation and deactivation of the source neurons located at macro- and micro-anatomically different brain regions and contributing to the record. The source neurons receive inputs from peripheral sensory as well as central nervous system. Thus, the expressions of the EEG vary in association with sensory-motor and psycho-physio-cognitive states and anatomical connectivities among the brain areas \cite{Gustavo,vec,ton}. Our contention may be supported by the fact that experimental and theoretical studies and gross EEG analyses have assigned causal relationship between the EEG changes and sleep-wakefulness \cite{vec,Gen,Genn,Fer}. We understand that many issues remain yet unanswered possibly due to lack of detailed EEG analyses and we believe analyses of EEG using higher mathematical tools might be helpful. Therefore, in principle appropriate deconvolution and decoding of the EEG are likely to be correlated with spatial and temporal source localization, including in association with  cognitive states. Further, results of such analyses possibly would differentiate between diseased and healthy states and may characterise those states with better temporal resolution. We hypothesized that in principle, analyzing the conscious-state specific EEG in association with sleeping and waking might take us closer towards unravelling the functional dynamics of the neurons located in different brain regions and are responsible for those conscious states. \\

We analyzed cortical EEG recorded from the frontal (motor area) and occipital (visual area) areas of chronically prepared freely moving rat brains during wakefulness, NREMS and REMS vigilance states (VS), as well as during the transitions to and from these vigilance states, the vigilance state-transitions (VST). Using network theory measures we demonstrate switching of functional connectivity between the frontal and occipital regions is  associated with predominance of frequency in the EEG during VS and VST, suggesting gradual transition between cognitive states. Also, the recording and analysis of EEG from the above regions would be constructive because these cortical regions are associated with learning skills and memory consolidation, respectively \cite{ram,wal}. \\

\begin{figure}
\label{fig1}
\begin{center}
\includegraphics[height=12cm,width=17cm]{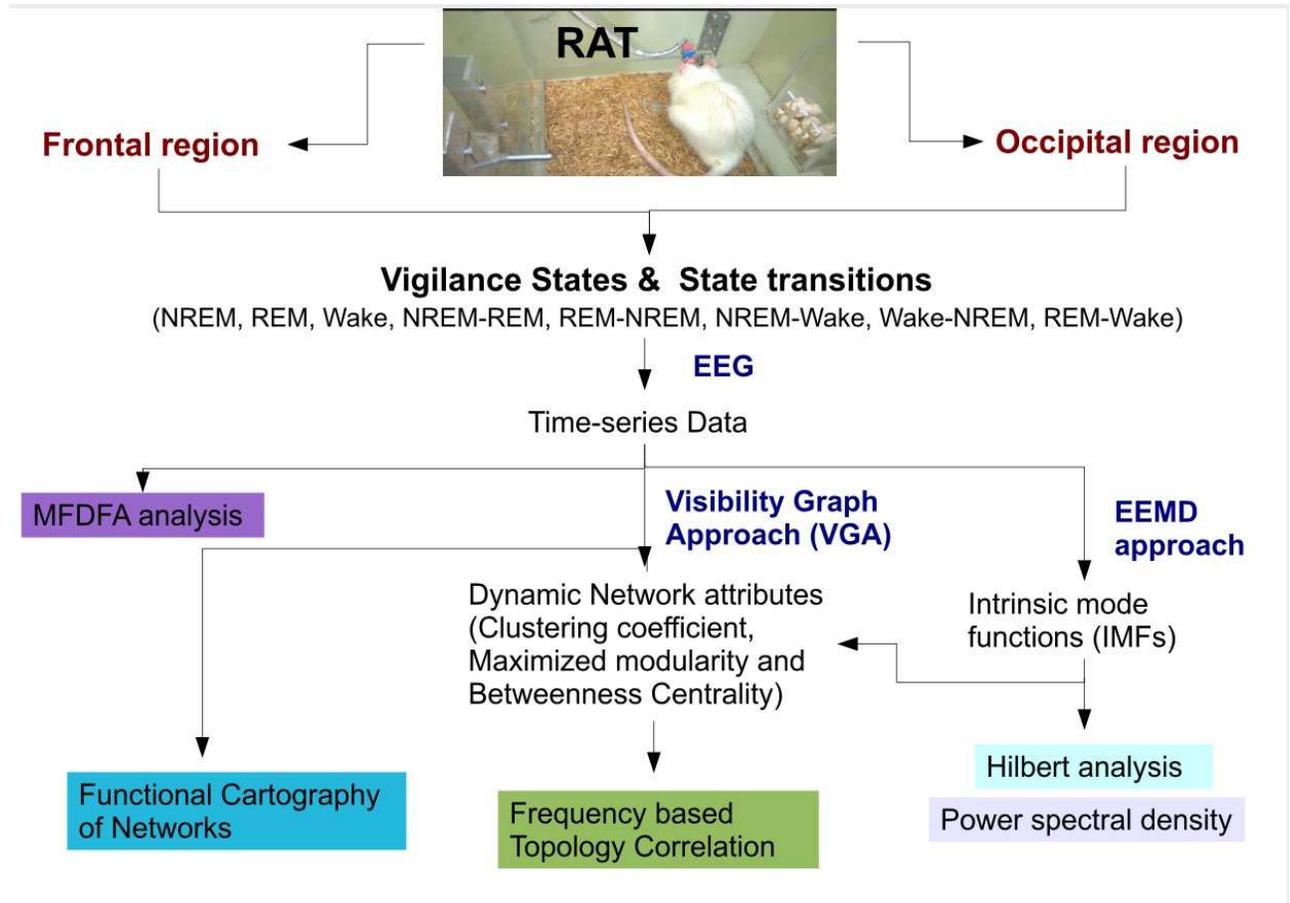}
\caption{\textbf{The flow chart demonstrates the workflow applied for the analysis.} }
\end{center}
\end{figure}

\begin{figure}
\label{fig2}
\begin{center}
\includegraphics[height=12cm,width=17cm]{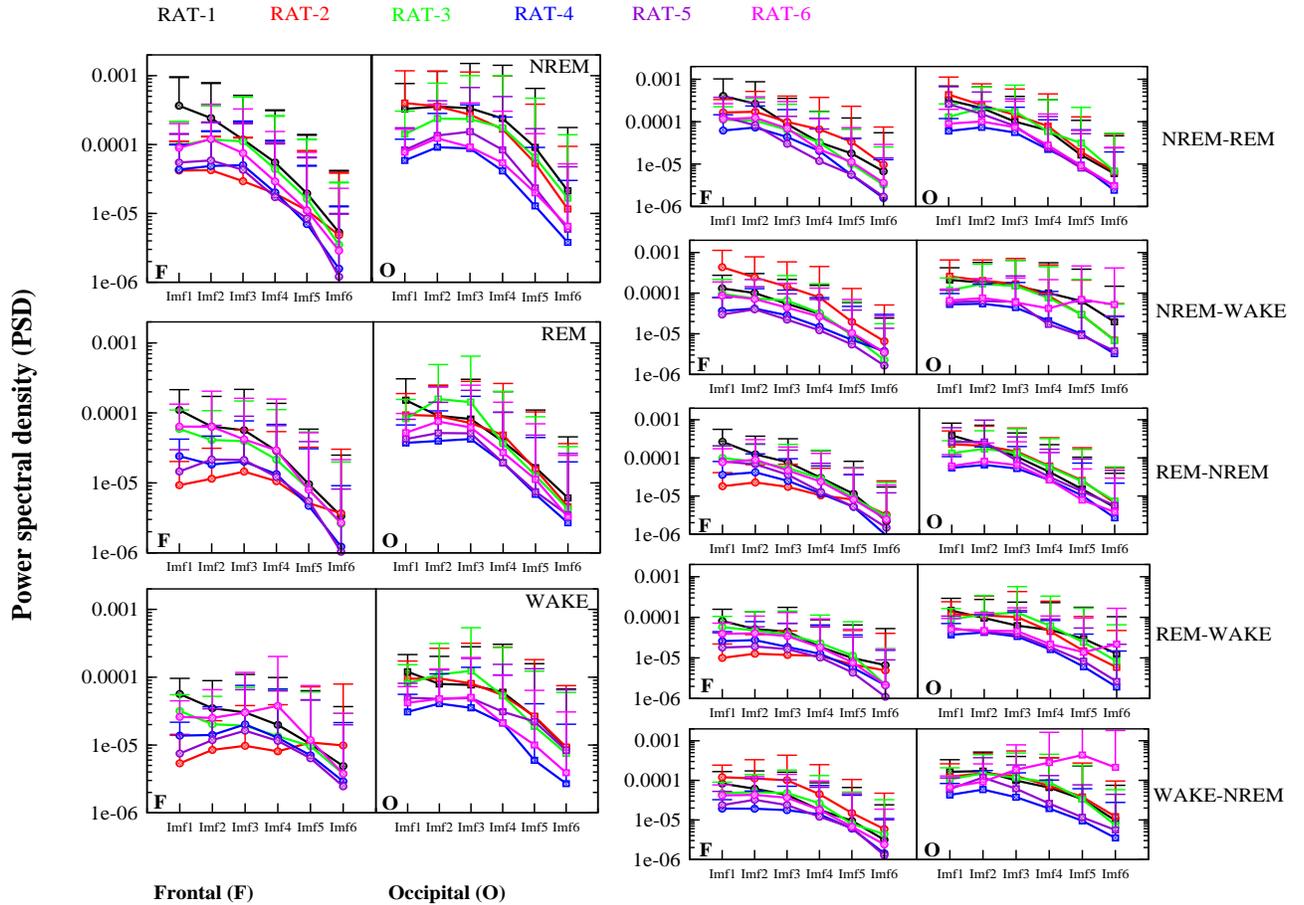}
\caption{\textbf{Power spectral analysis of rat frontal and occipital region.} \textbf{A)} The Average power spectral density (P) versus IMFs, for all vigilance states and state transitions of rat frontal (F) and occipital (O) region. The data for six rats is shown in different colors.}
\end{center}
\end{figure}

\begin{figure}
\label{fig3}
\begin{center}
\includegraphics[height=17cm,width=14cm]{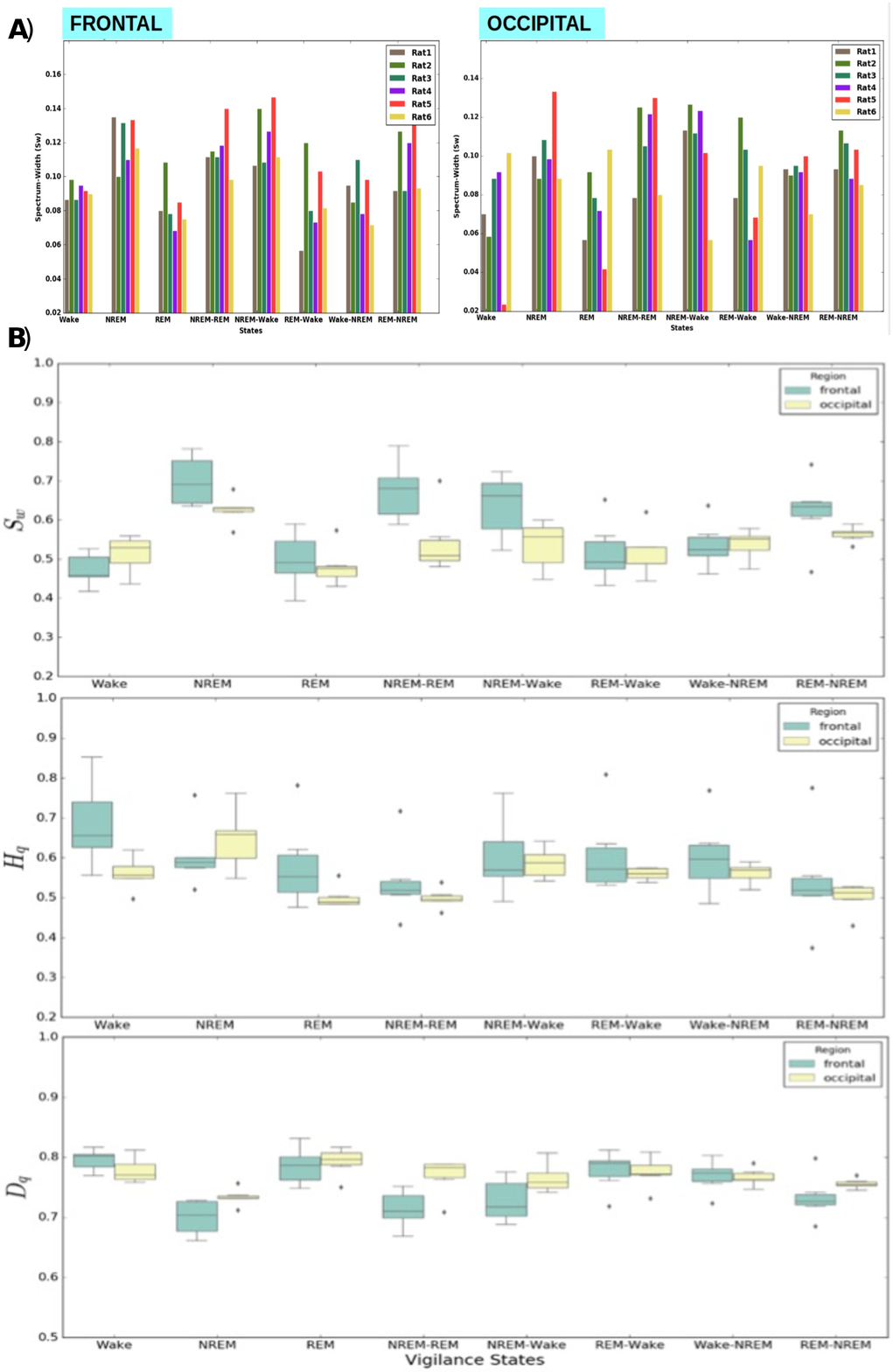}
\caption{\textbf{Multifractal analysis:} \textbf{A)} The multifractal spectrum width, $S_W$, for comparison of complexity among all states of frontal and occipital region.
\textbf{B)} The multifractal spectrum parameters $S_w$, $H_q$ and $D_q$ have been compared over frontal and occipital region at each vigilance state and state transition for the averaged data of six rats. Note- The points outside the error bars represent the rat outliers.}
\end{center}
\end{figure}

\begin{figure}
\label{fig4}
\begin{center}
\includegraphics[height=12cm,width=17cm]{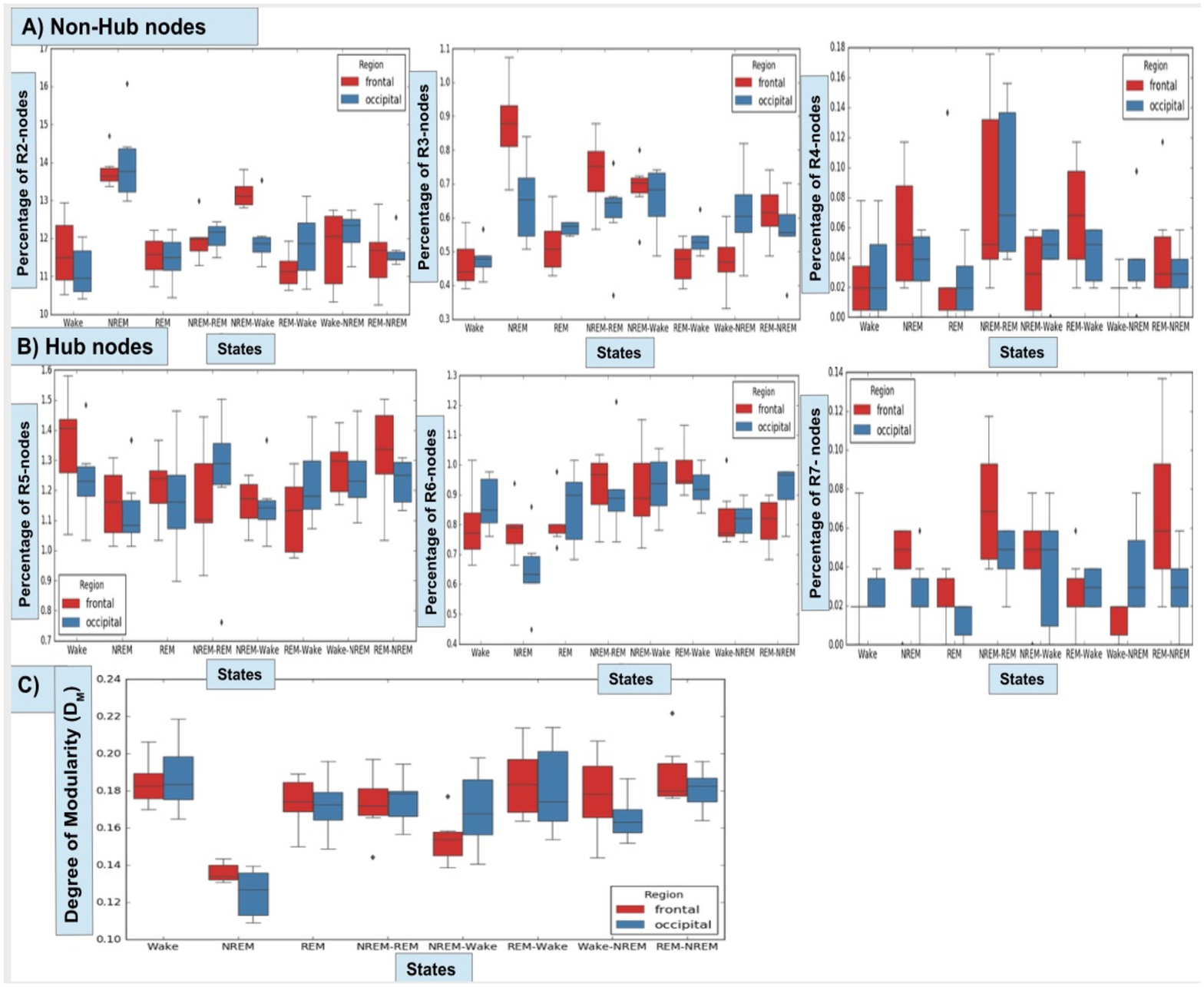}
\caption{\textbf{Degree of Modular organization in the frontal and occipital region:} Comparison of all vigilance states and state transitions among the frontal and occipital region, based on the A) percentage of R2, R3 and R4 non-hub nodes, B) percentage of R5, R6 and R7 hub nodes, and C) Degree of Modular-organization, $D_M$. The plots are showing values of six epochs (grey points as outliers) which is averaged over four rats. }
\end{center}
\end{figure}

\begin{figure}
\label{fig5}
\begin{center}
\includegraphics[height=18cm,width=18cm]{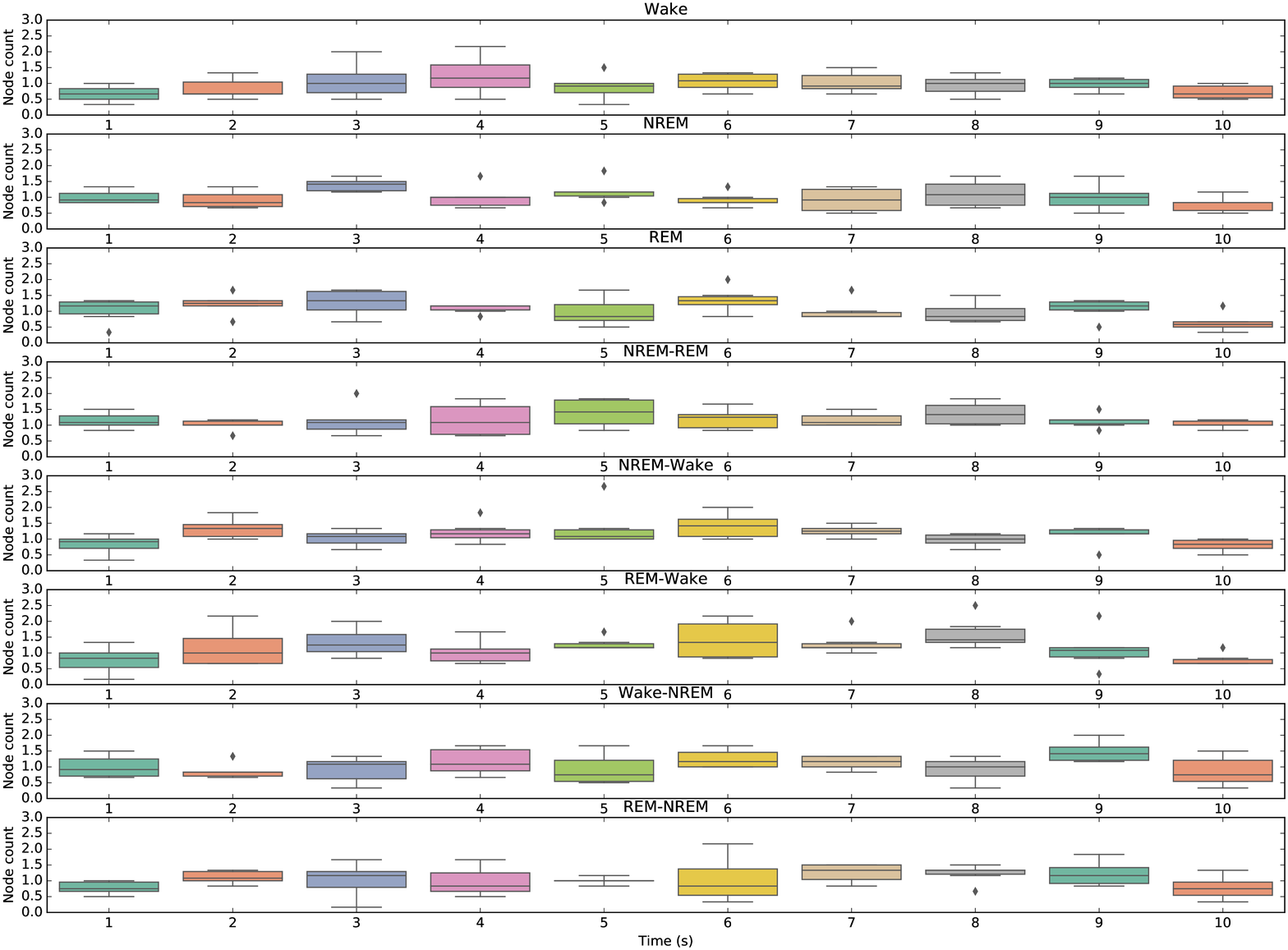}
\caption{\textbf{Average count of connector hubs:} The figure represents the average count of R6 hub nodes during each vigilance state and their state transitions with respect to time in the frontal region for the data of six rats. Note- The points outside the error bars represent rat outliers. }
\end{center}
\end{figure}

\begin{figure}
\label{fig6}
\begin{center}
\includegraphics[height=18cm,width=18cm]{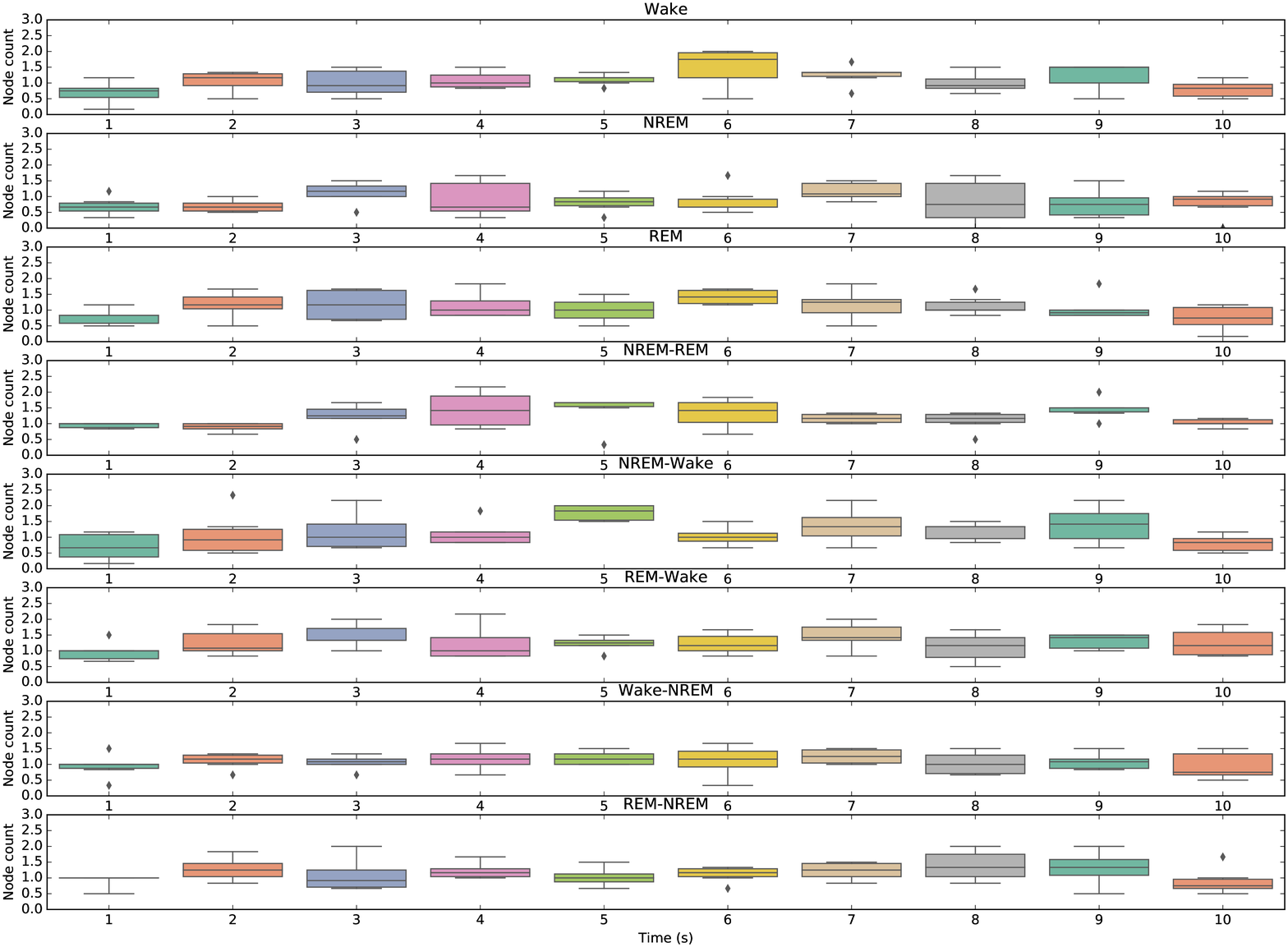}
\caption{\textbf{Average count of connector hubs:} The figure represents the average count of R6 hub nodes during each vigilance state and their state transitions with respect to time in the occipital region for the data of six rats. Note- The points outside the error bars represent rat outliers. }
\end{center}
\end{figure}

\begin{figure}
\label{fig7}
\begin{center}
\includegraphics[height=12cm,width=17cm]{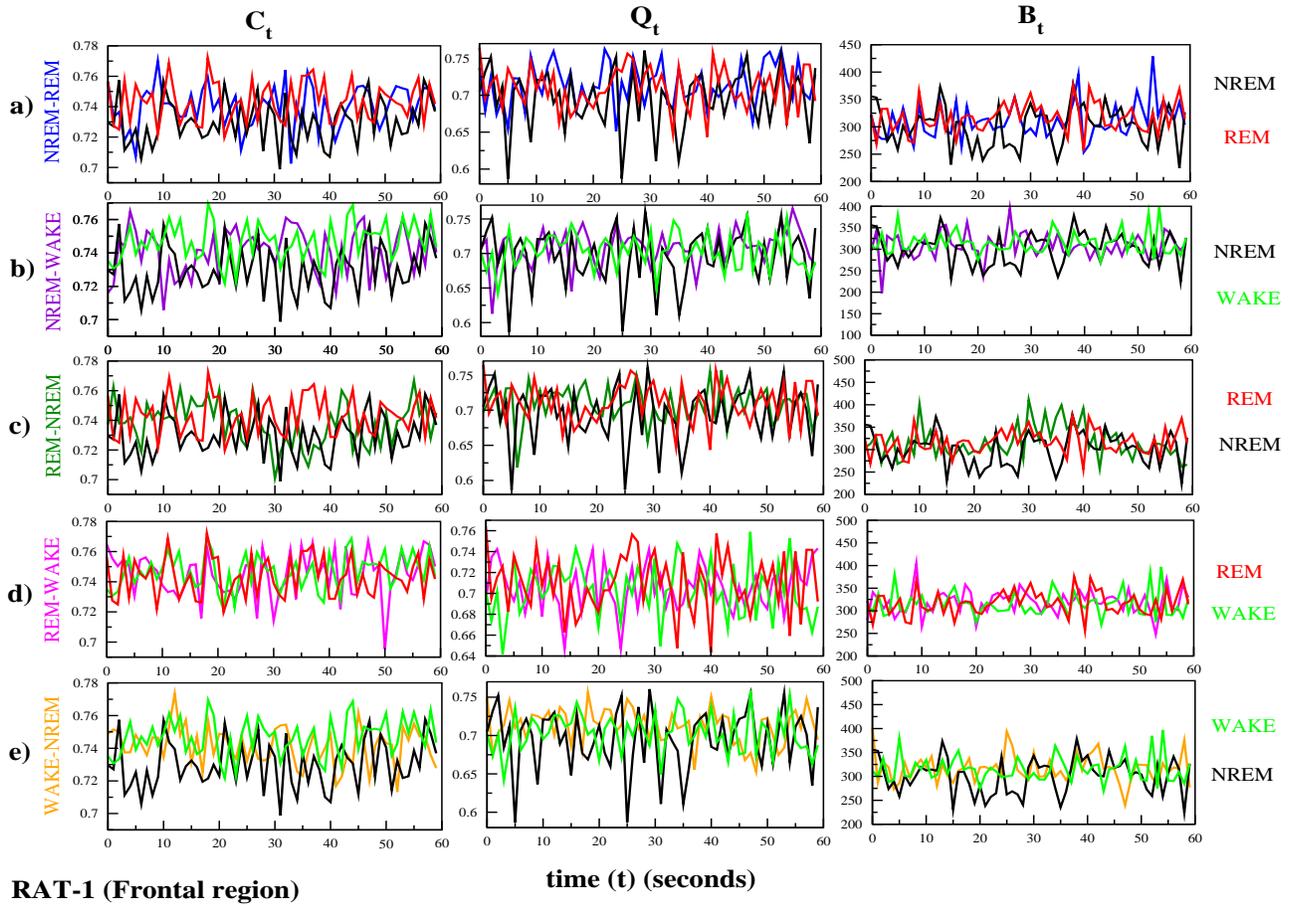}
\caption{\textbf{Characterization of dynamic network connectivity in the frontal region:} The first column plots the Clustering coefficient, $C_t$, the second column is the maximized Modularity, $Q_t$, and the third column is the Betweenness centrality, $B_t$, with respect to time $t$ for all the vigilance state transitions (mentioned in the left) compared with its constituting vigilance states. }
\end{center}
\end{figure}

\begin{figure}
\label{fig8}
\begin{center}
\includegraphics[height=12cm,width=17cm]{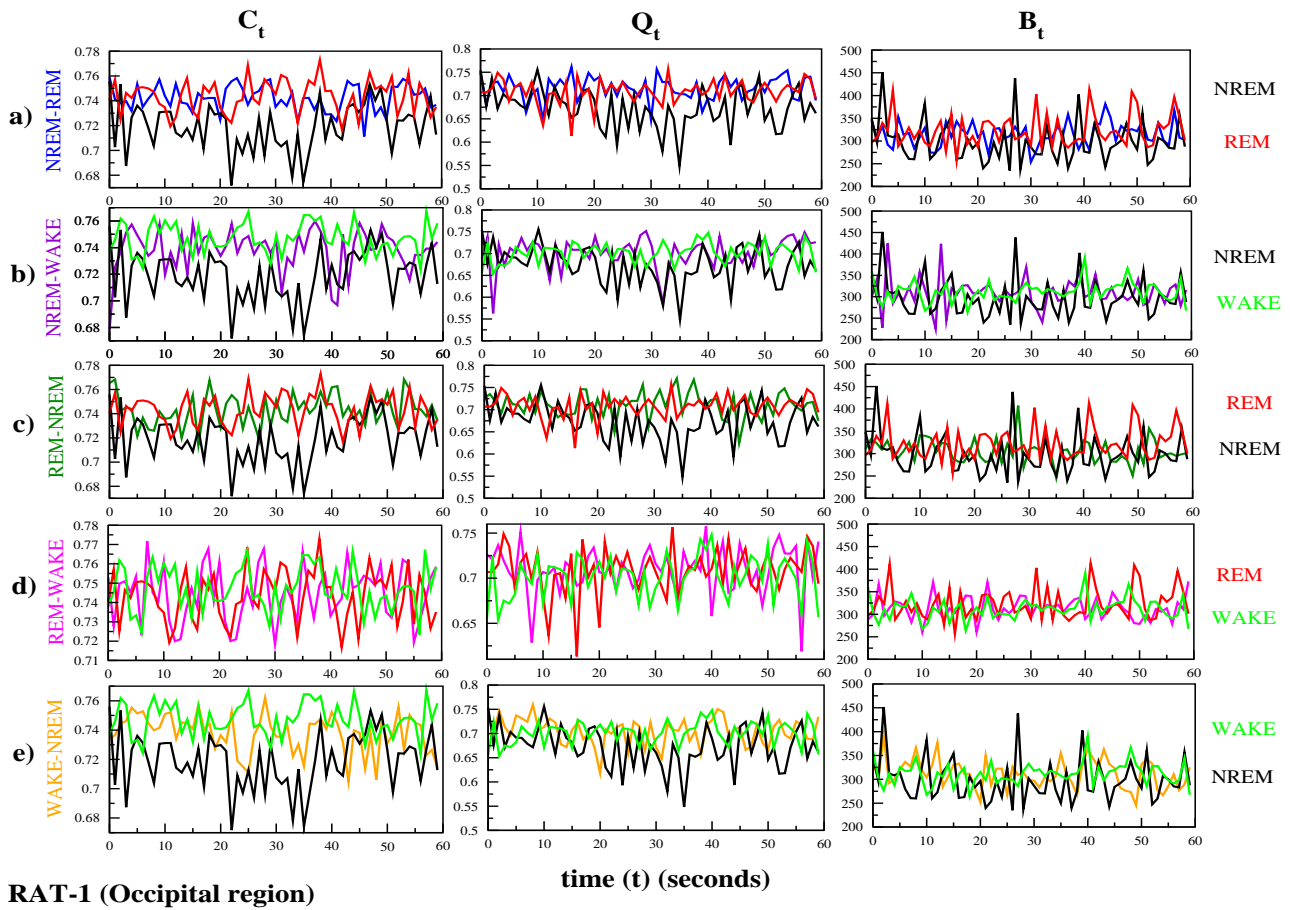}
\caption{\textbf{Characterization of dynamic network connectivity in the occipital region:} Similar to the above plot, the dynamic trend in $C_t$, $Q_t$ and $B_t$ as a function of time $t$.}
\end{center}
\end{figure}

\begin{figure}
\label{fig9}
\begin{center}
\includegraphics[height=14cm,width=17cm]{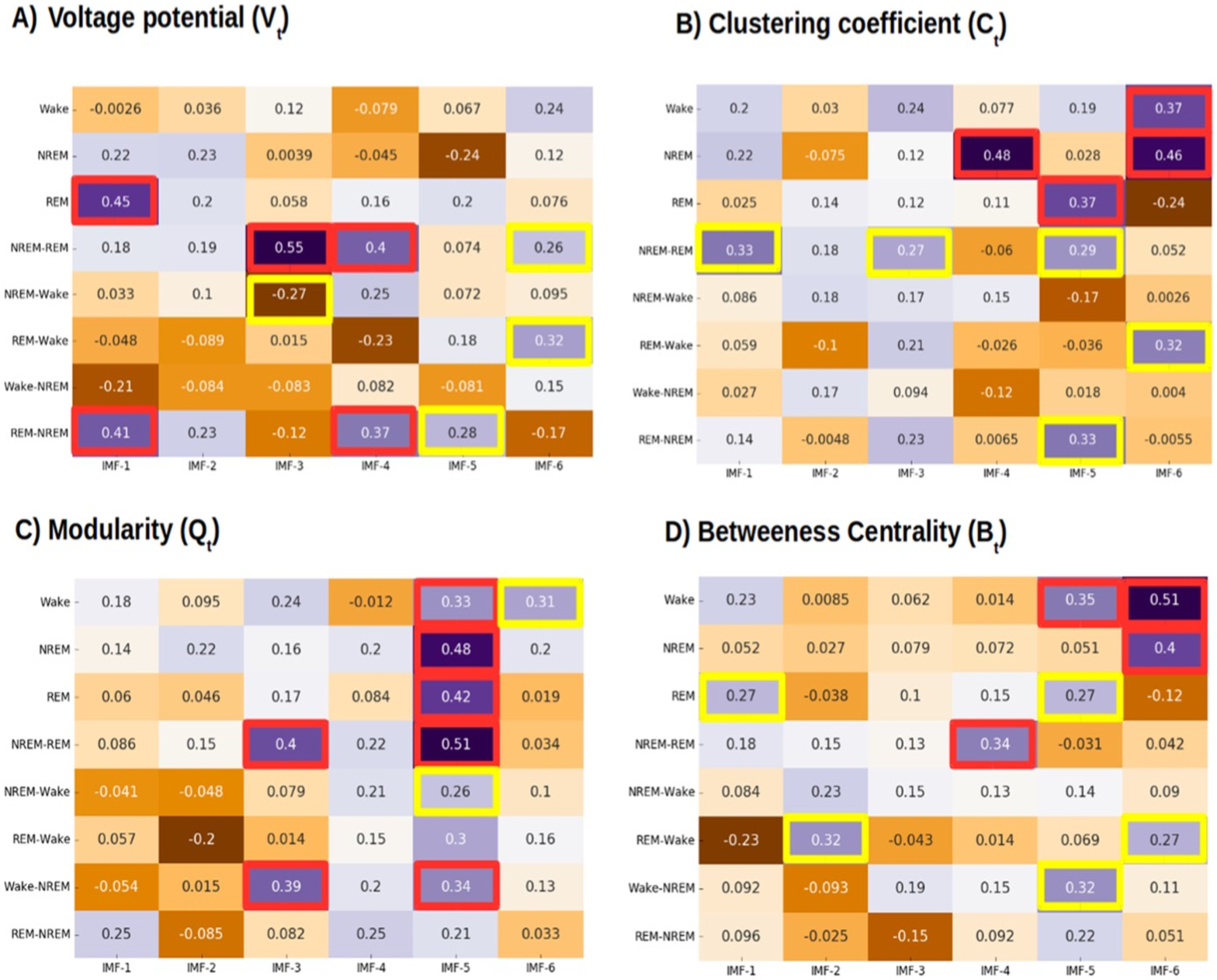}
\caption{\textbf{Topological Correlations among the frontal and occipital region:} It illustrates the Pearson correlation coefficient in the form of a heatmap for the A) Voltage potential, $V_{t}$, B) Clustering coefficient, $C_{t}$, C) Modularity, $Q_{t}$, and D) Betweeness centrality, $B_{t}$, of all the IMFs versus vigilance states and state transitions. This shows the data averaged over six rats. \textbf{Note-} The colored squares represents significant correlations with p-values less than 0.05 (yellow) and less than 0.01 (red). }
\end{center}
\end{figure}

\vskip 0.5cm
\noindent\textbf{\Large Results and Discussion}\\
We have carried out time-series analysis of the EEG as shown in the workflow (Fig.1). The flow-chart depicts the sequence of approaches we have implemented to analyse sleep-wake VS and VST on EEG recorded from the frontal and occipital cortex of the rat brain. We focused on 'Wake', 'NREMS' and 'REMS' as VS and defined five VST, namely, 'Wake-NREMS', 'NREMS-Wake', 'NREMS-REMS', 'REMS-NREMS' and 'REMS-Wake' (transition from Wake into REMS does not take place normally). The comparative analysis of these states and state transitions have been done along the following two paths: (a) Complexity analysis of the temporal signal through fractal characterization \cite{eke}, and, b) Computing the \textit{dynamic network attributes} and their topological correlations for signal oscillations. We collected EEG from six rats and analyzed six epochs (considered as trials) of 10s each from the 3 VS and 5 VST of each rat (For details see Methods).\\

\noindent\textbf{Signal decomposition and spectral analysis}\\
We performed the signal decomposition through the ensemble empirical mode decomposition (EEMD) \cite{wu} to attain constituting frequency oscillations as intrinsic mode functions or IMFs (See Methods). The instantaneous frequency of the IMFs 2-6, as obtained through hilbert transform, falls in the bandwidth of Gamma, Beta, Alpha, Theta, delta oscillations, respectively, (See supplementary Fig.1). The EEG signals with positive signal-to-noise ratio (SNR) were subjected to EEMD to further reduce noise. Neuronal activities can be characterized by the power spectral density (PSD), which is defined as the energy spectral density per unit time, of the temporal EEG in a state \cite{tsa,bjo}. In Fig.2, we have demonstrated average spectral density P (with error bars) for each IMF during each VS and VST of the frontal and occipital region. The power of each IMF quantifies the role of each oscillation in each state that acts as signatures for the state transitions in the brain dynamics. Since these are average values, the characteristic frequencies of states are not prominent; however, we have observed high spectral density in the occipital than in the frontal region. \\

\noindent\textbf{Complexity of Vigilance states and transitions}\\
Multi-fractal detrended fluctuation analysis (MFDFA) approach has been widely used over non-stationary temporal data to ensure the fractal behavior of dynamic patterns in complex biological systems \cite{zor,eke}.
The statistical fractals generated through physiological time series data show self affinity in terms of different scaling with respect to direction. The scaling parameters to define multi-fractal nature are fluctuation function $F_q$, hurst exponent $H_q$, mass exponent $t_q$, scaling exponent $h_q$ and fractal dimension $D_q$ (See Methods). The spectrum of multiple scaling exponents shows multifractal nature of the temporal signal in REMS, NREMS and Wake states (See supplementary Fig.2). However, to distinguish the extent of multi-fractal nature, we have defined a parameter called spectrum width, $S_w$, to measure the degree of complexity of the signal in a state-specific manner. The plots in Fig.3A shows $S_w$ (averaged over 6 trials) for each VS and VST in the frontal and occipital region of the rats. We have also plotted, averaged values over all rats, $S_w$, $H_q$ and $D_q$ (See Methods), as shown in Fig.3B, to show comparison between frontal and occipital region considering each VS and VST. The $H_q$ characterizes the average fractal structure whereas the $S_w$ mentions the deviation from the average fractal structure \cite{ihl}. Thus, the highly deviated structure would be considered as more complex or in certain terms less stable. The comparative analysis of these parameters (Fig.3B) suggests NREMS to be more complex state as compared to the other states. The NREMS is the first and can be considered the most dynamic state between wake-sleep transition \cite{olb,moru}. We argue that more complexity of the NREMS stage is due to the efforts put in to synchronize the neuronal activities from different regions in the brain (as reflected in the EEG) as opposed to desynchronized EEG during the other states \cite{sch}. Another notable observation is almost consistently high complexity of vigilance state transitions. This suggests state transitions to be a distinguished and transient complex state, that may require additional complexity generators to accomplish the process of transition. \\

\noindent\textbf{Modular Organization also gets affected during state transitions}\\
The functional brain networks are known to be organized in a modular manner \cite{meu}. Therefore, we thought of analyzing the level of modular organization in our vigilance states and state transitions. The modular organization of networks is characterized by the presence of hub nodes and non-hub nodes that constitute a functional module. A modular structure implies more within module connectivity than inter-connectivity among modules. We have followed a node classification based on the participation coefficient, \textit{$P_{i}$} and within-module degree or Z-score, \textit{$Z_{s}$}, values of nodes in the network, as per the study done in \cite{gui}. The categorization has labelled nodes as provincial (R5), connector (R6) and kinless (R7) module hubs and ultra-peripheral (R1), peripheral (R2), connector (R3) and kinless(R4) non-hubs. We evaluated the percentage of each category of the hub and non-hub nodes for all VS and VST among the frontal and occipital region and compared them to estimate their global and local functional connectivity trends. The Fig.4 shows comparison of the degree of modular organizations in the frontal and occipital region based on A) the percentage of R2, R3 and R4 non-hub nodes, B) percentage of R5, R6 and R7 hub nodes, and C) Ratio of hub/non-hub nodes,  during all VS and VST. The NREMS state outperforms wake and REMS with a high percentage of the non-hub nodes but lags behind them with low percentage of the hub nodes, and also depicts a low hub/non-hub ratio (as shown in Fig.4C). These findings characterize NREMS with more local and globally less activity. We observed that the ratio of hub/non-hub nodes for REMS is equivalent to wake state, as reported earlier \cite{and}. Another important observation is the degree of modular organization among frontal and occipital region is paradoxical during NREMS-WAKE and WAKE-NREMS state transitions. In the former occipital has a higher number, while in the later, the frontal region has a higher number of hub nodes. There is also evidence from past research that demonstrates REMS regulation through switching behaviour of REM-ON and REM-OFF neurons \cite{mall,kum}. Thus, we anticipate activation/inactivation of hub nodes as a mechanism of getting sleep from arousal state and vice-versa. \\
We have further characterized the connector hub (R6) that mostly forms inter-modular interactions, which plays a major role in determining the basis of the functional architecture of a network. We think that the activation/inactivation of R6 hubs would be important in the maintenance of a particular state. We have characterized switching (activation/inactivation) behaviour in each VS and VST over time in a comparative manner (Fig.5 and 6). The variation in the average count of connector hubs over different states would determine their activation/inactivation as key events during different states and state transitions. The switching mechanism of connector hubs at each time point, thus, signify their role in maintenance and transition of particular states through their functional interactions in the network. \\

\noindent\textbf{Dynamic network attributes and properties}\\ 
We generated the underlying network connectivity of the temporal data through visibility graph approach \cite{lac} for each time second (using data-points in that time instant). The network theory approach has been widely used to critically analyze complex brain networks \cite{bul}. The network attributes such as clustering coefficient (C), maximized modularity (Q) and betweeness centrality (B) are known to characterize network topology \cite{new1}. We measured the averaged values of network attributes for the 128-node network (as sampling frequency = 128Hz) at each time point $t$ in seconds and addressed them as \textit{dynamic network attributes}. We took six epochs (trials) of 10s duration from each VS and VST of the frontal and occipital region and continuously augmented their values to make a big time-series for each rat. We did so because state transition epochs cannot be obtained in chronological order. The \textit{dynamic network attributes}, thus, exhibit fluctuations in the functional connectivity of the underlying brain dynamics for each state and state transition \cite{dim}. We have shown in Fig.7 and Fig.8, the temporal variation trend in the average clustering coefficient, $C_{t}$, maximized modularity, $Q_{t}$, and betweeness centrality, $B_{t}$ during the state transitions and compared them with their {primary} vigilance states, for the frontal and occipital region,respectively, for one representative rat. The dynamic trend of these \textit{dynamic network attributes} during transition states mostly remained in between their respective vigilance states. This suggests the existence of state-transitions as an intermediary transient state. \\

\noindent\textbf{Frequency-based topology correlation}\\
The aforementioned approach of \textit{dynamic network attributes} has now been performed with the data from all the rats for their significant IMFs (first 6 IMFs) to obtain the frequency based temporal distribution of the topological attributes. {We, then, computed Pearson correlation coefficient from the temporal data of \textit{dynamic network attributes} among the frontal and occipital region, for each VS and VST. We also calculated the voltage potential, in order to compare with topological correlations, by simply averaging the data points at each time instant $t$ second. We have illustrated the correlation in network topology of frontal and occipital region during the vigilance states and state transitions, in terms of Clustering coefficient, $C_{t}$, Modularity, $Q_{t}$, and Betweeness centrality, $B_{t}$ in Fig. 9B, 9C and 9D, respectively and correlation in Voltage potential, $V_{t}$, in Fig.9A. The heat maps show the correlation values for all the significant IMFs (frequencies), defining interesting insights about their predominant role in maintaining coordination among the frontal and occipital region during respective state. The high correlation values marked with coloured squares are significant correlations with p-values either $<0.05$ (yellow square) or $<0.01$ (red square). These topological correlations exhibits synchronized behaviour, established between two regions, in a better way than simply correlating voltage potential.}   \\

\vskip 1cm
\noindent\textbf{\Large Methods}\\

\noindent\textbf{Animals}\\
Seven healthy, inbred male Wistar rats (250–300g) were used in the study. All the rats had free access to rodent food and water ab libitum and were maintained in 12:12 light: dark cycle at 25±1ºC ambient temperature. All experimental protocols were approved by the Institutional Animal Ethics Committee of Jawaharlal Nehru University, India. The principal guidelines for care and use of laboratory animals issued by the National Institutes of Health were followed. All steps were taken to minimize avoidable pain and discomfort to the experimental animals while conducting the experiments.

\noindent\textbf{Surgery}\\
Under isoflurane (Baxter, USA) induced surgical anaesthesia the rats were implanted with bipolar EEG, EMG and EOG electrodes for chronic sleep-wake recording, as reported earlier in \cite{sin}. During the surgical procedure body temperature, respiratory rate, and peripheral reflexes were monitored as a reflection of depth of anaesthesia. An incision was given on the skin, scalp was removed to clean and expose the skull. Pairs of stainless steel screw electrodes were fixed in the frontal (AP= +2 mm, 2 mm lateral) and occipital bones (AP=-5.8 mm, 3 mm lateral) of the skull, respectively \cite{pax}. Another screw electrode was fixed on the mid-line over the frontal sinus to serve as animal ground. Flexible insulated (except at the tip) wires were connected bilaterally into the external canthus muscles of the eyes for recording bipolar EOG and on either sides of the dorsal neck muscles for recording bipolar EMG. The free ends of these electrodes were soldered to a 9-pin female plug, which was then fixed onto the skull with dental acrylic. The rats were recovered with adequate standard post-operative care. The rats were acclimatized with the semi-sound proof recording chamber (Faraday cage) and wires from the 5th post-recovery day onwards. 

\noindent\textbf{Recording and scoring of data}\\
The EEG, EOG and EMG were recorded for 8hrs between 9:30AM to 6:30PM on chart paper using GRASS model 7H polygraph, as well as using Vital Recorder software (Kissei Comtech, Japan); the recording channels were calibrated before the recording. The signals were sampled at 128 Hz, with band-pass filter 0.1Hz to 100Hz for EEG and EMG and 0.3Hz to 30Hz for EOG. A notch filter was applied to remove the 50Hz cycle (line noise) during signal acquisition. The EEG recordings from surgically implanted electrodes have been recorded during the light phase of 8 hours from 10:00 to 18:00. EEG, EOG and EMG signals were manually analysed offline into wakefulness, NREMS and REMS using Sleep Sign software (Kissei Comtech, Japan) in bins of 5sec as described earlier \cite{sin,mal}.\\ 

EEG desynchronization, high muscle tone and frequent eye movements were considered wakefulness; epochs characterized by synchronized EEG, low muscle tone and negligible eye movements were considered NREMS, while REMS was characterized by desynchronized EEG, simultaneous muscle atonia in the EMG and rapid eye movements in the EOG following NREMS. The following criteria were applied to characterise transition states. It was marked when the period of an ongoing stage progressively declined, and the beginning of characteristic signals that represents either of the states, e.g. NREMS, REMS, or wake, appeared. The appearance of this state was confirmed by matching its characteristic criteria (EEG, EMG, and EOG), as mentioned above to define the vigilant state epochs of sleep-wake cycle. For example, the transition epoch for the Wake to NREMS was selected when the low amplitude EEG gradually preceded with the beginning of visually prominent high amplitude stable EEG waves. The signals shows EEG disappearance of wake and the generation of prominent synchronized waves of NREMS was marked for the analysis of wake to NREMS transition. Similarily, NREMS to REMS, REMS to NREMS, REMS to Wake and NREMS to Wake transition state epochs were selected. Such epochs of 10s duration were extracted from the sleep-wake transition cycle. \\

\noindent\textbf{Ensemble Empirical Mode Decomposition (EEMD)}\\
The EEMD generates a set of intrinsic mode functions (IMFs), adaptive to the nature of the signal, and aid in the reduction of noise in the signal \cite{wu,wu1}. This method disintegrates the signal into unique frequency components while keeping the signal in the time domain, therefore preferred over Fourier transform. We specified the number of siftings=10 and ensemble size=500 (see Methods). The signal time-series has been decomposed into 9 IMF's excluding the aperiodic residual (see Supplementary Fig.1A). These IMFs constitutes a frequency band specific to each IMF as computed through their Hilbert spectrum profile. We selected first six IMFs and observed that IMF-1 contained all the higher frequencies and noise of the signal whereas the IMF-2 to IMF-6 possessed frequencies in the range of brain oscillations $\delta$, $\theta$, $\alpha$, $\beta$ and $\gamma$, respectively, Supplementary Fig.1B. We did not consider the low-frequency oscillations ($<1$Hz) or the last few IMF's. We considered the first six IMF's for each epoch of the VS and VST, for the further analysis, in the frontal and occipital region of six rats. \\
This approach is a modified variant of the empirical mode decomposition
(EMD) method, used to decompose the signal into intrinsic mode functions
(IMFs) while maintaining the time domain \cite{hua}. The IMf's are
basically intrinsic oscillatory modes that comprise of different frequency
components present in the signal. This approach calculates ensemble
of trials by adding white noise of finite amplitude for each IMF component,
as computed through the basic EMD algorithm, and takes mean over ensembles
to remove the noise part. This has been shown to be a better method to
get rid of signal noise and it also resolves problem of mode mixing \cite{wu}.
The basic algorithm of EMD is as follows:\\
\begin{itemize}
\item In the temporal signal, S(t), we first identified all the local extrema i.e. maxima and minima. 
\item Then, connecting all the maxima and minima extremes with natural
cubic spline lines would form the upper, $e_{U}(t)$, and lower, $e_{L}(t)$,
envelopes. 
\item Computed the mean of the envelopes as $e_{M}(t)= \left[e_{U}(t)+e_{L}(t)\right]/2$. 
\item Deducted the envelope mean from the signal and label the difference as the proto-IMF, $h(t) = S(t)-e_{M}(t)$ . 
\item Checked the proto-IMF for the stopping criterion to determine if it is as per the definition of IMF. 
\item If the proto-IMF did not satisfy the definition, repeated steps (a) to (e) on h(t) as many times as needed till it attains the structure of an IMF. 
\item If the proto-IMF does satisfy the definition, assign the proto-IMF as an IMF component, $c_{j}(t)$. 
\item Repeat the operation step (a) to (g) on the residue, $r(t) = S(t)-c_{j}(t)$, as the data. 
\item The operation ends when the residue $r_{n}(t)$ contains not more than one extremum.
\end{itemize}

Thus, the signal S(t) is decomposed in terms of IMFs, $c_{j}(t)$,
as,
\begin{eqnarray}
S(t)=\sum_{j=1}^{n}c_{j}(t)+r_{n}(t)
\end{eqnarray}
where $r_{n}(t)$ is the residue of data $S(t)$, after n number of IMFs
are extracted. The IMFs being simple oscillatory functions with vacillating
amplitude and frequency is characterized with the following properties: \\

1. The number of local extrema and the number of zero-crossings must
either be equal or differ at most by one throughout the length of
a single IMF.

2. The mean value of the envelope defined by the local maxima and
the local minima should be zero at any data point.\\

\noindent\textbf{Hilbert spectrum}\\
{\noindent}Once intrinsic mode function (IMF) components are obtained, one can apply Hilbert transform to each IMF components to compute instantaneous frequency $\omega=\frac{d\theta}{dt}$. Then the original data can be expressed as,
\begin{eqnarray}
S(t)&=&RP\sum_{k=1}^{n}A_k(t)e^{i\int\omega_k(t)dt}~~~~~(Hilbert~ transform)\nonumber\\
&=&RP\sum_{k=1}^{n}A_k(t)e^{i\omega_k(t)},~~~~~(Fourier~ representation)
\end{eqnarray}
where, $A_k(t)$ is the amplitude of the signal which can be related to the energy associated with the signal. Even though each IMF represents generalized Fourier expansion, $A_k$ and $\omega_k$ are variables for nonstationary data, where, Hilbert transform gives us time dependent of these variables $A_k(t)$ and $\omega_k(t)$ which is expected \cite{Huang1}. This allows to represent each Fourier component by rectangular blocks with thickness $d\omega$ in the frequency-time $(\omega,t)$ space. Since energy density ($H$) associated with IMF component is the square of the amplitude $H\propto |A_k|^2$, one can get information of the energy associated with each IMF from the frequency-energy space projected from these blocks in ($\omega,t$) space. Hence, the energy distribution in frequency-time space, which is given by the the curves in frequency-energy-time space, is known as $Hilbert$ $spectrum$. Further, existence of energy at a particular $\omega$ reveals the possibility of appearing of local wave of that 
frequency $\omega$ \cite{Huang2}. The degree of stationarity, which is the measure of stationarity in the nonstationary time series data, can be measured by \cite{Huang3},
\begin{eqnarray}
\Gamma_{d}(\omega)=\frac{1}{T}\int_{0}^{T}\left[1-T\frac{H(\omega,t)}{h(\omega)}\right]^2d\omega
\end{eqnarray}
where, $h(\omega)=\int_{0}^{T}H(\omega,t)dt$ is the marginal spectrum, which is the measure of the total energy contributed from each frequency value. It also denotes the cumulated energy over the entire span of the data. Then, instantaneous energy of the IMF data can calculated by,
\begin{eqnarray}
E_I(t)=\int_{\omega}H^2(\omega,t)d\omega
\end{eqnarray}

\noindent\textbf{Power spectral density}\\
{\noindent}We used Burg method \cite{Proakis} to calculate power spectral density (PSD), which is a parametric method used to surmount spectral leakage effects that can not be cured by non-parametric method \cite{Subha}. The method, which is based on minimization of backward and forward prediction errors, calculates PSD of the signal $S(t)$ by,
\begin{eqnarray}
P(\omega)=\frac{E_k}{\left|1+\sum_{t=1}^{k}A_k(t)e^{-i\omega t}\right|^2}
\end{eqnarray}
where, $E_k$ is the total least square error which can be expressed as, $E_k=E_k^b +E_k^f$, where, $E_k^b$ and $E_k^f$ are backward and forward prediction errors respectively. This method is useful specially for EEG data analysis because it provides stable auto regressive model, useful estimation for short data, and close real value of the measurement. \\

\noindent\textbf{Multifractal analysis}\\
 The multifractal detrended fluctuation analysis (MFDFA) approach can be characterized as a measure of complexity and non-linearity in the complex signals such as EEG \cite{zor,eke}. The presence of multi-fractal nature (multiple scaling exponents), has been examined in the EEG time series using the DFA method, proposed by Kantelhardt et al. \cite{kant}, and implying it’s Matlab formulation as given by Espen A. F. Ihlen \cite{ihl}. Important parameters characterizing multifractality are scaling function (F), Hurst exponent (H), mass exponent (t), singularity exponent (h) and singularity Dimension (D). We computed above parameters to ensure the multi-fractal nature of the complex EEG signal. We have shown these parameters, in Supplementary Fig.2, for the trial dataset of one rat during the wake, REM and NREM states. The hurst exponent in range $0.5-1$ signifies a long-range correlated structure of the time series \cite{ihl}. 
 For a time series signal x\textbf{$_{j}$}of finite length \textit{l}\textbf{ }with random walk like structure, can be computed by the Root mean square (RMS) variation, $X_{i}=\sum_{j=1}^{i}\left(x_{j}-\langle x\rangle\right)$, where, $\langle x\rangle$ is the mean value of the signal, and i = 1,2, ..., \textit{l.} The signal X has been divided into $n_{s}=int(\frac{l}{s})$ non-overlapping segments of equal size \textit{s.} To avoid left-over short segments at the end, the counting has been done from both sides
therefore $2n_{s}$segments are taken into account. This defines the scale (s) to estimate the local fluctuations in the time series. Thus, the overall RMS, F, for multiple scales can be computed using the
equation, 
\begin{eqnarray}
F^{2}(s,v)=\frac{1}{s}\sum_{i=1}^{s}\left\{ X[(v-1)s+i]-x_{v}(i)\right\} ^{2}
\end{eqnarray}
where, $v$ = 1,2, ..., $n_{s}$ and $x_{v}(i)$ is the fitting trend in each segment $v$. The q-order RMS fluctuation function further determines the impact of scale (s) with large (+ q's) and small (-q's) fluctuations, as follows,
\begin{eqnarray}
F_{q}(s)=\left\{\frac{1}{2n_{s}}\sum_{v=1}^{2n_{s}}\left[F^{2}(v,s)\right]^{\frac{q}{2}}\right\}^{\frac{1}{q}}
\end{eqnarray}
The $q$-dependent fluctuation function $F_{q}(s)$ for each scale (s)
will quantify the scaling behaviour of the fluctuation function for
each $q$,
\begin{eqnarray}
F_{q}(s)\sim s^{H_{q}}
\end{eqnarray}
where, $H_{q}$ is the generalized Hurst exponent, one of the parameter
that characterizes multi-fractality through small and large fluctuations
( negative and positive q's) in the time series. The $H_{q}$ is related
to $q$-order mass exponent $t_{q}$ as follows,
\begin{eqnarray}
t_{q}=qH_{q}-1
\end{eqnarray}
From $t_{q}$ , the singularity exponent $h_{q}$ and Dimension $D_{q}$
is defined as,
\begin{eqnarray}
h_{q}=\frac{dt_{q}}{dq},~~~~~ D_{q}=qh_{q}-t_{q}
\end{eqnarray}
The plot of $D_{q}$ versus $h_{q}$ represents the multifractal spectrum
of the time series. The arc of the spectrum determines the complexity of the signal in terms of the range of scaling exponents i.e. defined as spectrum width $S_w$ \\

\noindent\textbf{Functional network connectivity }\\
We studied the underlying network topology and dynamics by applying the visibility graph algorithm \cite{lac}, on the temporal signals and respective IMFs corresponding to each VS and VST.
\textbf{Visibility graph method-} In this approach the number of data points
in the time-series would represent the number of neurons i.e. nodes
in the network. The connections between two neurons with data value
$n_{a}$ and $n_{b}$at time point $t_{a}$and $t_{b}$respectively
would be defined if the third neuron $n_{c}(t_{c})$ satisfies the
following condition,
\begin{eqnarray}
\frac{n_{b}-n_{c}}{t_{b}-t_{c}}>\frac{n_{b}-n_{a}}{t_{b}-t_{a}}
\end{eqnarray}
The extracted network would be connected, at least to its neighbors,
undirected and invariant according to the algorithm. The characteristic
properties of the time-series get delineated in the form of resultant
network. \\

\noindent\textbf{Network theory attributes}\\
Network theory is the much applied approach by researchers for characterizing
complex brain networks \cite{spo}. We have computed network theory
attributes at each time instant to study the variation in topological
organization over time. We choose basic network attributes that defines
the topology of a network, as follows, \\

\noindent\textbf{Clustering coefficient} (C) is a measure of degree that tells
completeness of a node's neighborhood . It characterizes how strongly
an ith node in the network is connected to the rest of the nodes and
can be estimated as the ratio of the number of its connected neighborhood
edges to the total number of edges possible, of a particular degree
$k_{i}$, $C(k_{i})=\frac{E_{i}}{k_{i}C_{2}}$, where, $E_{i}$ and
$k_{i}$ are the number of connected pairs of nearest-neighbors of
ith node \cite{new1}. We computed the dynamic clustering coefficient,
$C_{t}$, as average clustering coefficient of the network over each
time instant $t$ in seconds.\\

\noindent\textbf{Betweenness centrality} (B) measures the extent a node w is
traversed in the path of connecting node i and j via the shortest
path and is given by, $B(w)=\underset{i,j\,i\neq j\neq w}{\sum}\frac{d_{ij}(w)}{d_{ij}}$, where, $d_{ij}(w)$ is the number of geodesic paths from node i to
j d traversing through w, and $d_{ij}$ indicates total number of
geodesic paths from node i to j \cite{free}. It characterizes the
amount of information traffic diffusing from each node to every other
nodes in the network \cite{bor}. \\

\noindent\textbf{Modularity (Q)} The complex network structure can be forbidden
into communities or modules, specified with less than expected number
of connections among them \cite{new2}. To create a significant division
of a network the benefit function called modularity (Q) is defined
as, 
\begin{eqnarray}
Q&=&[Number\:of\:connections\:within\:community]-[Expected\:number\:of\:such\:connections]\nonumber\\
Q&=&\frac{1}{W}\sum_{[i,j=1]}^{N} [A_{ij}-B_{ij}]\delta_{c_{i},c_{j}},
\end{eqnarray}
The modularity Q, is maximized for good partitioning of the graph
G(S,E) with N as total nodes. The $A_{ij}$ and $B_{ij}$ defines
the exact and expected number of connections between nodes i and j.
$W=\underset{i,j}{\sum}A_{ij}$, $c_{i}$and $c_{j}$are the communities
nodes i and j belongs to,$\delta_{c_{i},c_{j}}$equals to 1, when
nodes i and j falls in same community and 0, if they do not. \\

\noindent\textbf{Functional cartography of networks}\\
The hierarchical-modular organization of brain networks is based on the presence of various high-degree nodes known as hubs, that follows the scale free topology. We can determine the topological distribution of hub and non-hub nodes in the network, as suggested in \cite{gui}, with their z-score being $Z_{s}\geq2.5$ and $Z_{s}<2.5$, respectively. The second-level segregation depends on their $P_{i}$ values that describes nodes as, R1 called ultra-peripheral non-hubs with all the edge connections in the same module $(P_{i} < 0.05)$; R2 called peripheral non-hubs with mostly intra-modular edges $(0.05 < P_{i} < 0.62)$; R3 called connector non-hubs with many inter-modular edges $(0.62 < P_{i} < 0.80)$; R4 called kinless non-hubs with homogeneous sharing of connections among modules $(P_{i} > 0.80)$; R5 called provincial hubs with most of intra-modular connections $(P_{i} < 0.30)$; R6 called connector hubs with majority of inter-modular associations $(0.30 < P_{i} < 0.75)$ and R7 called kinless hubs with homogeneous associations among all the modules $(P_{i} > 0.75)$ \cite{gui}.

{\noindent} The \textit{Participation coefficient} and \textit{Within module degree, $Z_i$} have been computed to categorize the network nodes as R1, R2, R3 and R4 non-hub nodes and R5, R6 and R7 hub nodes. The \textbf{Participation coefficient} ($P_{i}$) signifies the distribution of connections of a particular node i with respect to different communities
\cite{gui}. 
\begin{eqnarray}
P_{i}=1-\sum_{c=1}^{N_{m}}\left(\frac{k_{ic}}{k_{i}}\right)^{2},
\end{eqnarray}
where, $k_{ic}$ is the number of connections made by node i to nodes
in module c and $k_{i}$ is the total degree of node i. The $P_{i}$
value determines the distributional uniformity of the neuronal connections,
specified by the range 0-1. The escalating value signifies more homogeneous
allocation of links among all the modules.\\

{\noindent}The within-module degree or \textbf{Z-score} $(Z_{i})$ is another
measure to quantify the role of a particular node i in the module
$c_{i}$. High values of $Z_{i}$ indicate more intra-community connections
than inter-community and vice-versa \cite{gui}.
\begin{eqnarray}
Z_{i}=\frac{k_{i}-\bar{k_{c_{i}}}}{\sigma_{k_{c_{i}}}},
\end{eqnarray}
where, $k_{i}$ is the number of connections of the node i to other
nodes in its module $c_{i}$. $\bar{k_{c_{i}}}$ is the average k
over all nodes in the module $c_{i}$ and $\sigma_{k_{c_{i}}}$is
the standard deviation of k in $c_{i}$. The Brain connectivity toolbox has been used to compute the above mentioned network topological measures \cite{rub}.\\

{\noindent}We have defined the \textbf{Degree of modular organization, $D_{M}$}
of a network as the ratio of hub/non-hub nodes based on the classification
done in \cite{gui}. Also, we calculated the percentage of each R
node in order to get the idea of type and weightage of nodes (connections). 
\begin{eqnarray}
Percentage\,of\,R^{*}nodes&=&\frac{Number\,of\,R^{*}nodes}{Total\,number\,of\,nodes\,in\,the\,network}\\
D_{M}&=&\frac{Number\,of\,hub\,nodes\,(R5+R6+R7)}{Number\,of\,non-hub\,nodes\,(R1+R2+R3+R4)}
\end{eqnarray}
{*} It represents type of node.\\
$(S_{W})$, a measure of singularity exponents in the system.

\vskip 0.5cm
\noindent\textbf{\Large Summary}\\
Brain functional topology exhibits multifractal nature because of various functionalities and emergence of complexity and non-linearity in the neuron dynamics \cite{Gustavo,zor,shyam}. {The complexity of sleeping brain is undoubtedly proven and this study has done thorough analysis to investigate generators of complexity. This work extensively addresses the complex network connectivity in the frontal and occipital region of rat brain during the vigilance states of sleep, wake and among their state transitions. We have observed high power dominance in the Occipital than the frontal region (see Fig.2A). The multifractal analysis has shown significant change in the complexity of the VS and VST, as reported through the spectrum width (see Fig.3). Though it seems that NREM is highly complex state, through multifractal analysis, but the functional cartography of nodes has revealed low percentage of hub nodes in the NREM state (see Fig.4B) than REM and wake. This suggests NREM sleep to be functionally disconnected to some extent or globally less active state of brain. However, the high percentage of non-hub nodes (see Fig.4A) could be the reason of getting high non-linearity during NREM sleep. This makes sense to us that the brain is active at the local level but the activity is not getting transmitted at the global level, because of lesser number of functional hubs \cite{de}.} Another interesting observation is switching behavior in the number of activated hubs of frontal and occipital region during transition from NREM-Wake and Wake-NREM (see Fig.4C). We have also computed the fluctuations in the average count of connector hubs with respect to time (see Fig.5 and 6). This strongly anticipates the role of hubs as switches, that maintains dynamic functional complexity during vigilance states and also determines the transitions among states. \\
Further, we analyzed the role of different frequency oscillations in maintaining the topological correlations during the VS and VST. We have characterized the \textit{dynamic network attributes} for each IMF oscillation and their topological correlations among the frontal and occipital regions. Their correlation plots (see Fig.9) signify the role and contribution of an oscillation in maintaining coordination among frontal and occipital region during the particular vigilant states and state transitions. The dynamic altered connectivity trends during sleep stages and its comparison with wake has given us important insights regarding functional complexity of brain. \\
{\noindent}The extensive topological characterization done by us significantly classifies various brain states and state-transitions and portray hubs as markers of functional complexity. {Switching between hub-nodes for state transitions supports activation and deactivation of different sets of neurons for switching between states. However, the challenge is to understand the mechanism of such switchings during normal and diseased conditions.} As we had spatial limitations with the EEG data, so we did not specify the regional information of the hubs in a specific manner. However, this approach if applied to a whole brain data would give more fruitful insights. In future, we would appreciate further characterization and prediction of the state-transitions based on their complexity information.

\newpage
\appendix

\begin{figure}
\label{fig1}
\begin{center}
\includegraphics[height=16cm,width=17cm]{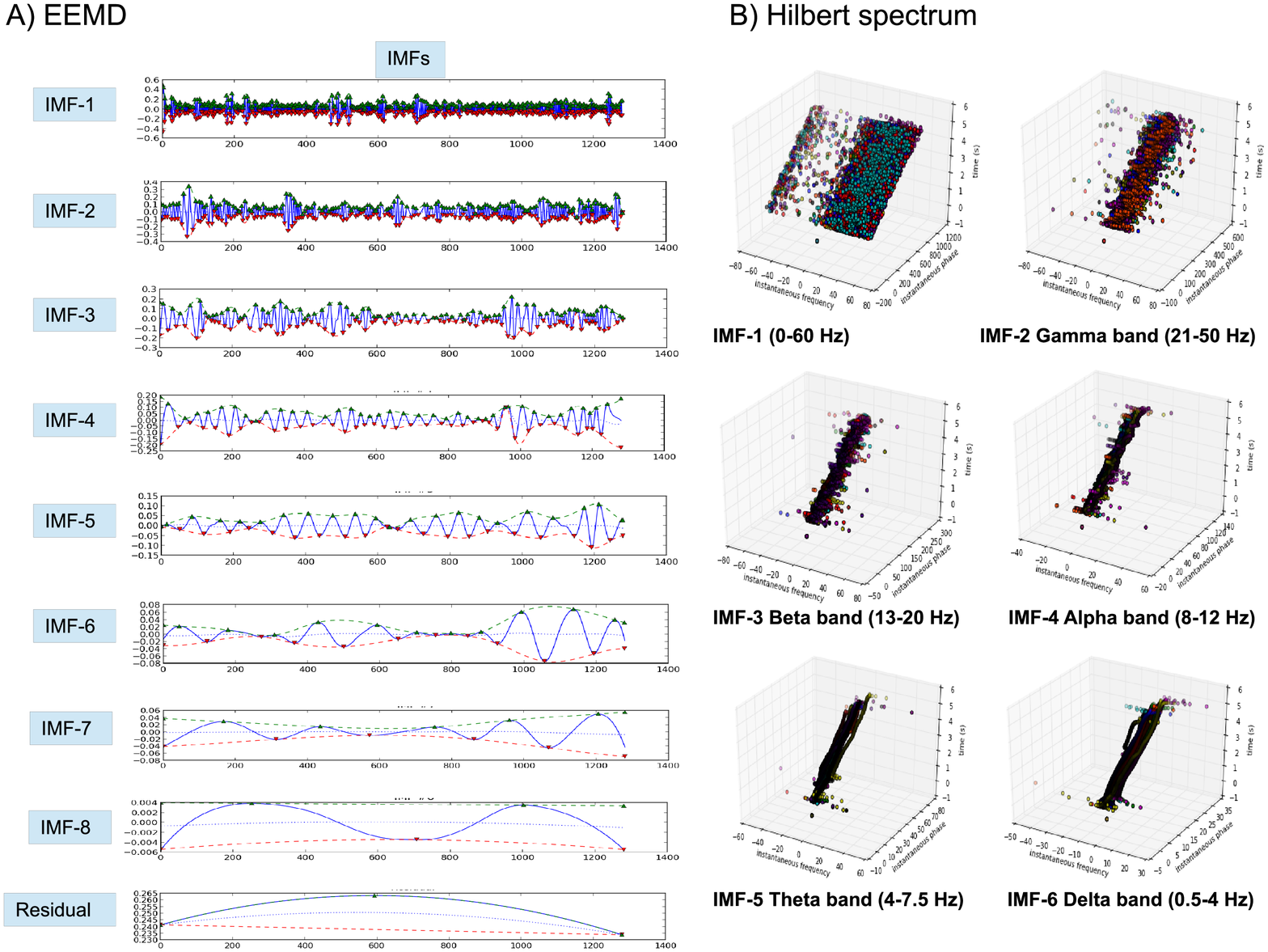}
\caption{\textbf{IMFs and Hilbert spectrum of EEG data:} A) constructed IMFs from the original experimental EEG data on rat; B) we have shown the plot of instantaneous frequency (x-axis) versus instantaneous phase (y-axis) and time (z-axis) for data of 12 epochs for a particular animal and state but this is a general observation for all states.} 
\end{center}
\end{figure}

\begin{figure}
\label{fig2}
\begin{center}
\includegraphics[height=12cm,width=17cm]{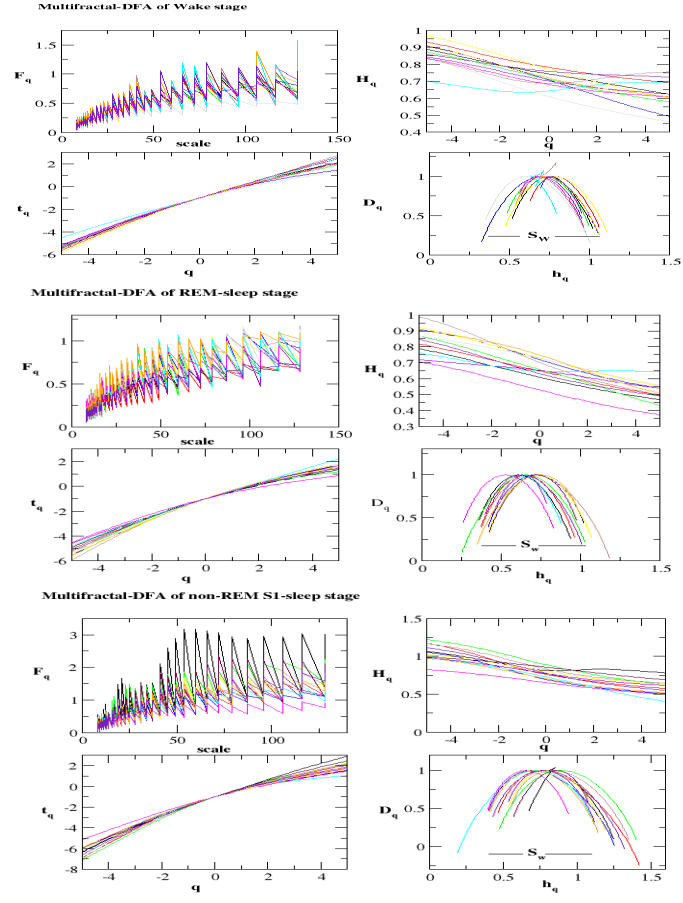}
\caption{\textbf{MF-DFA approach:} The plots of q order scaling parameters (clockwise from left) fluctuation function, $F_q$ versus scale (or segment length), the hurst exponent, $H_q$, the mass exponent, $t_q$, and the fractal dimension, $D_q$ versus singularity exponent, $h_q$, for the wake, NREM and REM states. This is the result of a trial dataset of 12 epochs to show consistency in trend. }
\end{center}
\end{figure}

\end{document}